\documentclass[structabstract]{aa}  
\usepackage{natbib} 
\bibpunct{(}{)}{;}{a}{}{,} % to follow the A&A style 
 
\usepackage{graphicx}
\usepackage{mathpazo}

\usepackage{amsmath, amssymb}
\begin{document}
   \title{Laboratory H$_2$O:CO$_2$ ice desorption data: entrapment dependencies and its parameterization with an extended three-phase model}
\titlerunning{laboratory H$_2$O:CO$_2$ ice desorption data}
   %\subtitle{}

   \author{Edith C. Fayolle
          \inst{1}
          \and
          Karin I. \"Oberg\inst{2}
          \and 
          Herma M. Cuppen
          \inst{1,3}
          \and
          Ruud Visser
          \inst{3}
          \and
          Harold Linnartz
          \inst{1}
          }

   \institute{Sackler Laboratory for Astrophysics, Leiden Observatory, Leiden University, PO Box 9513, 2300 RA Leiden, The Netherlands\\
              \email{fayolle@strw.leidenuniv.nl}
         \and
             Harvard-Smithsonian Center for Astrophysics, MS 42, 60 Garden Street, Cambridge, MA 02138, USA
          \and 
             Leiden Observatory, Leiden University, PO Box 9513, 2300 RA Leiden, The Netherlands
             }

   \date{}

% \abstract{}{}{}{}{} 
% 5 {} token are mandatory
 
  \abstract
   {Ice desorption affects the evolution of the gas-phase chemistry during the protostellar stage, and also determines the chemical composition of comets forming in circumstellar disks. From observations, most volatile species are found in H$_2$O-dominated ices. }
   { The aim of this study is first to experimentally determine how entrapment of volatiles in H$_2$O ice depends on ice thickness, mixture ratio and heating rate, and second, to introduce an extended three-phase model (gas, ice surface and ice mantle) to describe ice mixture desorption with a minimum number of free parameters. 
   }
   {Thermal H$_2$O:CO$_2$ ice desorption is investigated in temperature programmed desorption experiments of thin (10 -- 40~ML) ice mixtures under ultra-high vacuum conditions. Desorption is simultaneously monitored by mass spectrometry and reflection-absorption infrared spectroscopy. The H$_2$O:CO$_2$ experiments are complemented with selected H$_2$O:CO, and H$_2$O:CO$_2$:CO experiments. The results are modeled with rate equations that connect the gas, ice surface and ice mantle phases through surface desorption and mantle-surface diffusion.
    }
   {The fraction of trapped CO$_2$ increases with ice thickness (10--32~ML) and H$_2$O:CO$_2$ mixing ratio (5:1 -- 10:1), but not with one order of magnitude different heating rates.  The fraction of trapped CO$_2$ is 44 -- 84 \% with respect to the initial CO$_2$ content for the investigated experimental conditions. This is reproduced quantitatively %for binary ice mixtures 
   by the extended three-phase model that is introduced here. The H$_2$O:CO and H$_2$O:CO$_2$:CO experiments are consistent with the H$_2$O:CO$_2$ desorption trends, suggesting that the model can be used for other ice species found in the interstellar medium to significantly improve the parameterization of ice desorption.}
   {}

   \maketitle
%
%________________________________________________________________

\section{Introduction}
\label{sec:intro}

In pre-stellar cores, cold outer protostellar envelopes and protoplanetary disk midplanes, most molecules, except for H$_2$, are frozen out on dust grains, forming ice mantles. The main ice component in most lines of sight is H$_2$O, followed by CO and CO$_2$, with a typical abundance of $(0.5 - 1.5) \times 10^{-4}$ for H$_2$O ice with respect to H$_2$ around solar-type protostars \citep{Vandishoeck_06}. Infrared observations of pre-stellar cores show that most CO$_2$ ice and some of the CO ice is mixed with H$_2$O \citep{Knez_05}.  The remaining CO and CO$_2$ are found in separate ice layers. Based on these observations, H$_2$O and CO$_2$ are thought to form simultaneously on the grain surface during the early stage of cloud formation. When the cloud becomes denser, gas phase CO freezes out on top of the water-rich ice, resulting in a bi-layered ice mantle, as described in \cite{Pontoppidan_08}.
	
 Once the pre-stellar core starts collapsing into a protostar, it heats its environment, including the icy grains. This results in the desorption of the CO-rich layer into the gas phase, in structural changes in the water-rich ice layer, and eventually  in the desorption of the water-rich layer \citep{Pontoppidan_08}. Such an ice desorption scheme provides most of the gas phase reactants for the chemistry taking place at later stages in these warm regions \citep{Doty_04}. It is therefore crucial to understand ice mixture desorption and to effectively implement it in astrochemical networks. The aim of this study is to provide a laboratory basis for this process and to demonstrate how it can be modeled both in the laboratory and in space.
	
	Laboratory experiments have provided most of the current knowledge about ice thermal desorption, including desorption energies for most pure simple ices \citep{Sandford_88,Sandford_90, Fraser_01, Collings_04, Oberg_05, Brown_07, Brown_10}. Desorption from ice mixtures differs from pure ice desorption because of different binding energies between the mixture components (e.g., the CO binding energy increases from 830 K in pure ice to  $1180\, \rm K$ in H$_2$O-dominated ice mixtures \citep{Collings_03_b}) and because of trapping of volatile species in the H$_2$O hydrogen-bonding ices \citep{Collings_04}.  %In general, 
Volatile components therefore desorb from H$_2$O-rich ice mixtures at a minimum of two different temperatures, corresponding to the desorption of the species from the surface of the H$_2$O ice and from molecules trapped inside the bulk of the H$_2$O ice, which only start desorbing at the onset of H$_2$O desorption. Additional desorption is sometimes observed at the temperature for pure volatile ice desorption and during ice re-structuring, \textit{e.g.}, at the H$_2$O phase change from amorphous to crystalline \citep{Viti_04}. This H$_2$O restructuring occurs at $\sim$140 K in the laboratory (for astrophysical timescales the re-structuring temperature and desorption temperature decrease), which is close to the onset of H$_2$O desorption \citep{Collings_04}.
	
	Of the different ice mixture desorption features, the entrapment of volatile species in H$_2$O ice is astrochemically the most important to quantify. The trapping of CO in a water ice results in a factor of five increase in the effective desorption temperature. In a recent cloud core collapse model, this corresponds to trapped CO desorbing at 30 AU from the protostar compared to pure CO ice desorbing at 3000 AU. The case is less dramatic, but still significant, for CO$_2$, which desorbs at $\sim$300 AU when pure, and at 30 AU if trapped in H$_2$O ice \citep{Aikawa_08, Visser_09}. Efficient ice trapping may therefore allow some volatiles to stay frozen on the dust grains during accretion of envelope material onto the forming protoplanetary disk \citep{Visser_09}. 
 
There are only a few models that have 
incorporated the effects of ice mixture desorption. \cite{Collings_04} investigated the desorption of 16 astrophysically relevant species from H$_2$O:X 20:1 ice mixtures. \cite{Viti_04} and \cite{Visser_09} used the results of \cite{Collings_04} to split up the abundance of volatiles in up to four different flavors, with different desorption temperatures. These correspond to the fraction of each ice desorbing at the pure ice desorption temperature, from a H$_2$O surface, during H$_2$O ice restructuring and with H$_2$O, respectively. This approach has provided information on the potential importance of ice trapping for the chemical evolution during star formation. However, this model does not take into account specific ice characteristics such as ice thickness, volatile concentration and heating rate, on which the amount of trapped volatiles in the water ice may also depend \citep{Sandford_88}. These characteristics need to be determined experimentally to correctly parameterize step models, where such are sufficient to model ice desorption. Strong dependencies on e.g. ice thickness or concentration would however warrant the development of a more continuous parameterization of ice desorption than the assignment of flavors.

These dependencies are naturally included in a few ice mixture desorption models of specific binary ices \citep{ Collings_03_b, Bisschop_06}. The molecular specificity of these models, together with a large number of fitting parameters has, however, prevented their incorporation into larger astrochemical models. Therefore, in most gas-grain networks, desorption is still treated as if ices were pure, disregarding volatile entrapment in the water matrix \citep[e.g.][]{Aikawa_08}.
	
	Another problem with current gas-grain codes is that evaporation is often incorporated as a first-order process, while it is experimentally found to be a zeroth-order process with respect to the total ice abundance for ices thicker than one monolayer. Desorption models from the last decades have shown the necessity of using a zeroth order kinetics \citep{Fraser_01,Collings_03_b}. Incorporating ice desorption as a first-order process with respect to the total ice abundance effectively means that molecules throughout the whole ice are allowed to desorb at the same time, which is non-physical \citep{Fraser_01, Bisschop_06}. This can be solved by treating the bulk and surface of the ice as separate phases as it has been done by \citet{Pontoppidan_03} and \citet{Pontoppidan_08} for CO ice desorption and by \citet{Collings_05} for H$_2$O desorption and crystallization. Its successful use in astrochemical models makes this approach an attractive option to parameterize laboratory ice desorption, since the results can then be easily transferred into an astrophysical context. In this family of models, molecules are only allowed to desorb from the surface, which is continuously replenished by molecules coming from the mantle, and therefore the desorption kinetics are automatically treated correctly. This model results in a zeroth-order desorption behavior, in agreement with the experiments, since the number of molecules available for desorption remains constant in time. The model also results in trapping of volatiles in the bulk of the ice since the mantle molecules cannot desorb into the gas phase. The three-phase model we build on was first introduced by \cite{Hasegawa_93}, but despite its advantages in treating different ice processes, it has not been generally used for ice mixture desorption, nor has it been further developed, presumably because it did not correctly reproduce the experimentally observed amount of volatiles trapped in the water ice.
	
	The goals of the present study are first to experimentally characterize how the trapping efficiency of CO$_2$ in H$_2$O ice depends on different ice characteristics (with complementary experiments on CO and tertiary mixtures) and second to use these experiments as a guide to improve our understanding of the trapping process within the three-phase model framework. The description of the extended three-phase model is explained in Section \ref{sec:model}.  The experiments used to get information on the volatile entrapment and to calibrate the model are described in Section \ref{sec:exp}. Laboratory results on H$_2$O:CO$_2$ ices, complemented by H$_2$O:CO and H$_2$O:CO$_2$:CO ice desorption results, are presented in Section \ref{sec:res_exp}. Section \ref{sec:mod} presents the model fitting parameters and model results. Finally, the consequences of treating ice mixture desorption with the extended three-phase model under astrophysical conditions are discussed in Section \ref{sec:astro}.

\section{Desorption Model}

This study addresses the desorption of volatiles mixed with water and how to predict the fractions of volatiles in the ice and gas phase during a warm-up of the ice. The model is a system of rate equations based on the \cite{Hasegawa_93} model, but with the addition of diffusion. It aims at providing a solution for the amount of volatiles trapped in water with respect to the ice characteristics that can be directly included into astrochemical models, as used by \cite{Viti_04} and \cite{Visser_09}. The model applies to species in the water-rich ice layer; the interface with an upper CO-rich ice layer is not treated here.

\label{sec:model}

\subsection{Basic three-phase model}

 \begin{figure}[t]
  \centering
   \includegraphics[scale=0.5]{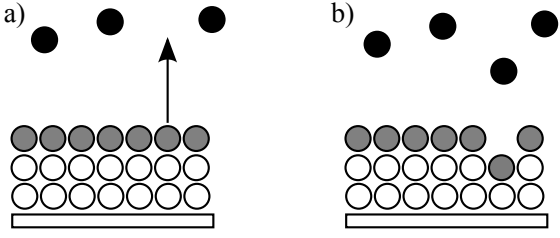}
      \caption{Cartoon defining the ice mantle (white), ice surface (gray), and gas phase (black) according to the \cite{Hasegawa_93} three-phase model. Panels a) and b) show the different phases before and after a desorption event.}
         \label{cartoon_model}
   \end{figure}

The model used here to predict the trapping of volatile species in a water dominated ice is based on the three-phase model by \cite{Hasegawa_93}. In this model, gas-grain interactions are addressed by considering three phases: the gas phase, the surface of the ice and the bulk/mantle of the ice (Fig. 1). The original model includes reactions between species in both gas and solid phase, as well as accretion from the gas to the ice and thermal and non-thermal desorption. Thermal desorption alone is presented here. The model is based on the principle that molecules can only desorb from the surface into the gas phase and that the mantle molecules can only migrate to the surface following the desorption of a surface molecule. The time-dependent gas abundance of species $i$ is given by

\begin{equation}
\frac{\mathrm{d}n^{\rm g}_i}{\mathrm{d}t} = R_{\rm evap}
\label{eqr_3}
\end{equation}

\noindent where,

\begin{equation}
R_{\rm evap} = \left(\nu \, \mathrm{ e}^{-E_i/T} \right) n^{\rm s}_i 
\label{eqr_evap}
\end{equation}

\noindent with $n^{\rm g}_i$ and $n^{\rm s}_i$ the gas phase and surface abundance of species $i$, respectively, $E_i$ its binding energy in K, and $\nu$ a pre-exponential factor taken equal to $10^{12}\rm~ s^{-1}$, which is a standard value for physisorbed species \citep{Biham_01}.
The surface abundance of species $i$ can be written as:

\begin{equation}
\frac{\mathrm{d}n^{\rm s}_i}{\mathrm{d}t} = - R_{\rm evap} + R_{\rm repl}
\label{eqr_1}
\end{equation}

\noindent where,

\begin{equation}
R_{\rm repl} = \alpha \left[\sum_j \left( \nu \, \mathrm{ e}^{-E_j/T} \right) n^{\mathrm{ s}}_j \right] \frac{n^{\mathrm{ m}}_i}{\sum\limits_{j} n^{\mathrm{ m}}_j}
\label{eqr_repl}
\end{equation}

\noindent where $n^{\rm m}_i$ is the mantle abundance of species $i$, $\sum n^{\rm m}_j$ the total number of molecules in the mantle and $\alpha$ is the ice coverage on the surface, which is set to 2~ML to account for surface roughness. The first term in Eq. \ref{eqr_1} represents the loss of molecules $i$ from the surface by thermal desorption. The second term is related to the replenishment of the surface sites by mantle molecules: the empty sites created by the desorption of any type of species from the surface, $\sum\limits_{j} \left(\nu \, \mathrm{ e}^{-E_j/T} \right) n^{\mathrm{ s}}_j$, are statistically filled by molecules coming from the mantle. The probability for these molecules to be species of type $i$ is equal to its mantle fraction, $\frac{n^{\mathrm{ m}}_i}{\sum n^{\mathrm{ m}}_j}$.
The mantle abundance, $n^{\rm m}_i$, of species $i$ changes according to

\begin{equation}
\frac{\mathrm{d}n^{\rm m}_i}{\mathrm{d}t} = - R_{\rm repl}.
\label{eqr_2}
\end{equation}

Because of the term $\frac{n^{\mathrm{ m}}_i}{\sum n^{\mathrm{ m}}_j}$, the replenishment of the surface phase by the mantle molecules during ice mixture desorption depends only on the mixing ratio of each species in this model, \textit{e.g.}, for a H$_2$O:CO$_2$ 1:1 ice mixture, a molecule that desorbs into the gas phase has a 50\% chance to be replaced by a water molecule and a 50\% chance to be replaced by a CO$_2$ molecule. This results in desorption of some volatile species around the pure ice desorption temperature and the rest remains trapped in the water ice since water molecules quickly saturate the surface phase.

The ice abundances $n^{\rm s}_i$ and $n^{\rm m}_i$ are all in cm$^{-3}$, a unit directly related to the gas phase abundance. The abundance of species $i$ on the surface is defined via the relation

\begin{equation}
n^{\rm s}_i=N^{\rm s}_i~ n^{\rm d}
\label{eqr_gassolid}
\end{equation}

\noindent where  $N^{\rm s}_i$ is the average number of molecules $i$ on the grain surface, and $n^{\rm d}$ is the dust abundance. The same relation applies for the mantle abundance.

\subsection{Extended three-phase model}

The original three-phase model does not account for the preferred replenishment of the surface phase by volatile mantle species or that volatile species may diffuse more easily in the ice compared to water. \cite{Oberg_09_c} showed that this diffusion can result in segregation of the ice components, which is important for temperatures well below the desorption energy of most volatile species in an ice mixture. This demixing mechanism changes the surface replenishment probabilities proposed in the original three-phase model by \cite{Hasegawa_93}. Our proposed extension of the three-phase model accounts for this by introducing a mantle-surface diffusion term. Trapping of volatiles still occurs, but the surface-mantle diffusion of volatiles is enhanced compared to the original model, resulting in that more than 50 \% of the empty sites are filled by volatiles species. Quantitatively, this changes the surface and mantle abundances $n^{\rm s}_i$ and $n^{\rm m}_i$ as follows:

\begin{equation}
\frac{\mathrm{d}n^{\rm s}_i}{\mathrm{d}t} = - R_{\rm evap} + R_{\rm repl} + R^{\rm diff}_i,
\label{eqr_4}
\end{equation}

\noindent and

\begin{equation}
\frac{\mathrm{d}n^{\rm m}_i}{\mathrm{d}t} = -R_{\rm repl} - R^{\rm diff}_i
\label{eqr_5}
\end{equation}

\noindent with

\begin{equation}
R^{\rm diff}_i = f_i \, \nu \, \left[ n^{\rm s}_{\rm H_2O} \, \frac{n^{\rm m}_i}{\sum\limits_{j} n^{\rm m}_j} \, \mathrm{e} ^{-E_{\rm diff} / T} - n^{\rm s}_i \,  \frac{n^{\rm m}_{\rm H_{2}O}}{\sum\limits_{j} n^{\rm m}_j} \, \mathrm{e}^{-E_{\rm diff} / T}\right],
\label{eqr_6}
\end{equation}

\noindent and
\begin{equation}
E_{\rm diff} = \left(\epsilon ^{\rm swap}_{\mathrm{H_2O} - i} - \frac{E_{\rm H_2O}-E_i}{2} \right)
\label{eq_energy}
\end{equation}

\noindent for $\textit{i} \neq \rm H_2O$, and where $\epsilon^{\rm swap}_{\mathrm{H_2O} - i}$ is the energy barrier for a volatile molecule $i$ and a water molecule to swap ( i.e. change position) within the ice  and $f_i$ a fraction between 0 and 1 that is described below. The expression for the gas phase abundance remains unchanged (see Eq. \ref{eqr_3}). The diffusion term $R^{\rm diff}_i$ is added to the surface abundance (subtracted from the mantle abundance) to enhance the mantle to surface circulation for a volatile species $i$ at the expense of the water, thus, $R^{\rm diff}_i$ is expressed differently for volatile and water molecules. 
The volatile ice diffusion rate depends on the balance of the probability of volatile molecules to move from the mantle to the surface at the expense of a water molecule and on the probability of the reverse process.
This swapping process probability depends on the energy barrier $\epsilon^{\rm swap}_{\mathrm{H_2O} - i}$ of the process and on the energy difference before and after the swap, equal to $E_{\rm H_2O}-E_i$. The diffusion rate for the water molecules is the negative sum of the diffusion rates for the volatiles, $R^{\rm diff}_{\rm H_2O} = - \sum_i R^{\rm diff}_i$, since the total abundance of molecules in the mantle and in the surface is not affected by the diffusion process. A similar formalism was used to describe H$_2$O:CO$_2$ segregation in \citet{Oberg_09_c}; an exchange of a surface H$_2$O molecule and a mantle volatile is generally energetically favorable because H$_2$O forms stronger bonds than volatile species and a mantle H$_2$O molecule can form more bonds compared to a surface H$_2$O molecule.

From segregation studies of binary ices, it has become clear that only a limited fraction of the mantle participates in the mantle--surface circulation and that this fraction depends on the initial ice mixture ratio \citep{Oberg_09_c}. This is represented by the fraction $f_i$ 

\begin{equation}
f_i = 1 - \frac{n_i^{\rm m,ini} - c_i (x_i^{\rm ini})^{\beta}}{n_i^{\rm m}}
\label{eqr_8}
\end{equation}

\noindent where $n_i^{\rm m,ini}$ is the number of mantle molecules $i$ initially in the ice, $c_i$ an empirical factor determined for each volatile $i$, and $x_i^{\rm ini}$ the initial mixing ratio of volatiles $i$ with respect to water. The expression $c_i (x_i^{\rm ini})^{\beta}$ describes the number of mantle molecules available for segregation for a particular ice mixture before the onset of desorption and follows the form found by \cite{Oberg_09_c} in ice segregation experiments when $\beta$ is set equal to 2. The term $n_i^{\rm m,ini} - c_i (x_i^{\rm ini})^{\beta}$ is the number of mantle molecules protected from segregation. When the later expression exceeds the current number of volatile mantle molecules $n_i^{\rm m}$, $f_i$ reaches zero and segregation stops, \textit{i.e.}, the diffusion of volatile mantle molecules to the surface stops. Thus this definition results in a gradual slowdown of the 'upward' mantle-surface diffusion of volatile species, regulating the trapping characteristics of H$_2$O ice for different volatiles.

We have tested the performance of this extended three-phase model on the desorption of mixed H$_2$O:CO$_2$ ices by comparing model and experimental TPD experiments, where the model TPDs are constructed using the rate equations \ref{eqr_3}, \ref{eqr_4} and \ref{eqr_5}. In the model TPDs the initial ice temperature is raised in steps proportional to the heating rate and at each time step the rate equations from the three-phase model are applied to calculate the temperature dependent desorption and diffusion rates. The desorption rate of the volatile is what is plotted in the TPD curves.

TPD experiments of pure ices are performed to determine the binding energies $E_i$. The other free parameters that are used to optimize the model are the swapping energies $\epsilon ^{\rm swap}_{\mathrm{H_2O} - i}$ between H$_2$O molecules and volatiles $i$ and the empirical factor $c_i$ used to parameterize the diffusion of volatiles $i$ from the mantle to the surface. These two parameters are determined by performing TPD experiments of binary ice mixtures of H$_2$O:CO$_2$ with different mixing ratios, thicknesses and heating rates and by comparing the output of the model with the experimental trends, \textit{i.e.}, the amount of volatile species that remains trapped in the water ice at temperatures higher than the desorption temperature of the volatile species. 

It is important to note that the model does not include the finite pumping speed during experiments. This will affect the derived desorption barriers and these are therefore not meant to replace the ones derived from more detailed pure ice experiments in the literature. As long as the pumping rate is constant with temperature, excluding the pumping rate will not affect the determined ice fraction that desorbs at a low temperatures versus the fraction that desorbs with H$_2$O. This is a reasonable assumption above the pure volatile ice desorption temperatures, where cryopumping is no longer efficient. The derivation of $c_i$ and $\epsilon ^{\rm swap}_{\mathrm{H_2O} - i}$ should therefore not be affected by this simplification.

\section{Experiments}
\label{sec:exp}

 \subsection{Experimental parameters}
 
The experiments in this study are chosen to simultaneously provide data directly relevant to ice desorption in different astrophysical environments (with different ices) and to construct a proof-of-concept model for ice mixture desorption.  The focus is on CO$_2$ desorption from H$_2$O ice mixtures, one of the most important ice systems around protostars, with supporting experiments on CO desorption. While interstellar ices are expected to be complex mixtures, it is still useful to investigate desorption from binary H$_2$O:volatile ice mixtures since the H$_2$O:volatile interactions are expected to dominate the desorption process in space, both because H$_2$O is the major ice constituent and because H$_2$O generally forms stronger bonds with itself and with volatiles than volatiles do. This hypothesis has been further tested by performing TPD experiments of two tertiary H$_2$O:CO$_2$:CO ice mixtures.

The ice thickness and structure in the experiments are chosen to be as similar as possible to the existing observational constraints on interstellar ices. Interstellar ices are estimated to be less than 100 monolayers (ML) thick from the maximum amount of oxygen available for ice formation. The experimentally grown ices are between 10 and 40 ML, since it is only possible to quantify ice thicknesses up to a certain limit (40 ML in our case) using reflection-absorption infrared spectroscopy \citep{Teolis_07}. Information on ice structure in space is limited, but the lack of a water dangling vibration at 3700 cm$^{-1}$ suggests a less porous ice than typically produced in the laboratory. We minimized the porosity of ice analogues by injecting gas perpendicularly to the cold surface when growing the ices \citep{Stevenson_99,Kimmel_01}.

\subsection{Experimental procedures}

All desorption experiments are performed with CRYOPAD. This set-up has been described in detail elsewhere \citep{Fuchs_06}. The set-up consists of an  ultra high vacuum (UHV) chamber with a base pressure of $\sim10^{-10} \, \rm mbar$ at room temperature. Ices are grown on a  gold-coated substrate situated at the center of the chamber that can be cooled down to $16\, \rm K$ by a close cycle He cryostat. The relative sample temperature is controlled with a precision of $ 0.1 \, \rm K$ using a resistive heating element and a temperature control unit. The absolute sample temperature is given with a $2\, \rm K$ uncertainty. The system temperature is monitored with two thermocouples, one mounted on the substrate face, the other on the heater element.

A fourier transform spectrometer is used for reflection-absorption infrared spectroscopy (FT-RAIRS) to record vibrational absorption signatures of molecules condensed on the gold surface. The spectrometer covers $700 - 4000 \, \rm cm^{-1}$ with a typical resolution of $1\, \rm cm^{-1}$ and an averaged spectrum consists of a total of 256 scans.  Ice evaporation is induced by linear heating of the substrate (and ice) in TPD experiments. RAIR spectra are acquired simultaneously to monitor the ice composition during the TPD.  A quadrupole mass spectrometer (QMS) is positioned at $4 ~ \rm cm$, facing the ice sample to continuously analyze the gas-phase composition mass-selectively and to obtain desorption curves of evaporating molecules during the TPD experiments.
 
Mixtures and pure gas samples are prepared from $^{13}$CO$_{2}$ (Indugas, min 99\% of $^{13}$C), CO$_{2}$ (Praxair, 99\% purity), $^{13}$CO (Cambridge Isotope Laboratories, 98\% purity) and from gaseous water at the saturation pressure of a de-ionized liquid sample at room temperature. The de-ionized water is purified by three freeze-pump-thaw cycles. The samples are prepared separately, then injected in the chamber via an inlet pipe directed along the normal of the gold surface. In all gas samples, an isotopologue of CO was used to separate the QMS signal from background CO and N$_2$. Similarly, an isotopologue of CO$_2$ was used to minimize the overlap in RAIR spectra between CO$_2$ ice and atmospheric CO$_2$ gas outside the UHV chamber. 

H$_2$O and CO$_2$ ice amounts are determined directly using the RAIRS band strengths provided by \cite{Oberg_09,Oberg_09_b} for CRYOPAD. From these measurements the absolute ice thicknesses are known within 50\%. The relative ice abundance uncertainties are smaller, $\sim$20\%, and due to small band strength variations with ice composition and temperature.
 
Table \ref{desorbexps} lists the set of TPD experiments performed to calibrate and test the desorption model presented in Section \ref{sec:model}. The TPD experiments begin with the deposition of pure or mixed ice samples on the gold substrate cooled to $16-19 \, \rm K$, and continue with a slow heating of the ices at a constant specified rate until the desorption of the molecules from the surface is complete. The evaporated gas phase molecules are continuously monitored by the QMS. RAIR spectra of the ices are acquired before heating to determine ice thicknesses and mixture ratios as described above. Spectra are also recorded during the warm-up as a second independent way to determine the ice composition and to monitor eventual structure modifications.

The infrared data are reduced by subtracting a local baseline around the molecular features. Mass spectrometric data are reduced by subtracting the ion current from species present in the background for each mass channel. Absolute yields cannot be directly obtained by the QMS since it is situated away from the ice sample ($4 ~ \rm cm$) and thus some of the desorbing molecules may get pumped away before detection. All QMS desorption rate curves are therefore normalized in such a way that the time-integrated desorption rate from the various species corresponds to their infrared spectrally measured ice abundance at the beginning of each experiment.

 \begin{table}[!h]
\centering
\caption{Overview of the desorption experiments}             % title of Table
\label{desorbexps} 
\begin{scriptsize}
\begin{tabular}{l l c c c c c}      
\hline\hline  
             % inserts double horizontal lines
Exp. & Sample     & Ratio  & Thick.    & Heat. rate  & \multicolumn{2}{c}{Trapped CO$_2$/CO ice \%}\\  
         &                 &            &    (ML)          &         (K.min$^{-1}$)  &   wrt. CO$_2$/CO &  wrt. H$_2$O  \\% & model\\  % table heading 
\hline                        % inserts single horizontal line

1      & H$_2$O   & -        &  24                       &       1                                      &                                 -  & -\\%& -\\
2               & $^{13}$CO$_2$   & -     &  6      &       1&-& -\\% &-         \\
3              & $^{13}$CO   & -     &  6     &      1        &- & -\\%&- \\
4             &H$_2$O:CO$_2$ & 10:1 & 12 & 1  &         62 & 6.2\\%&-\\
5            &H$_2$O:CO$_2$ & 10:1 & 19 & 1& 75& 7.5\\% & - \\
6             &H$_2$O:CO$_2$ & 10:1 & 32 & 1 & 84& 8.4\\%&-\\
7          &H$_2$O:$^{13}$CO$_2$ & 5:1 & 32 & 1&64& 12.8 \\%&-\\
8          &H$_2$O:$^{13}$CO$_2$ & 5:1 & 18 & 10&62& 12.4\\% &-\\
9          &H$_2$O:$^{13}$CO$_2$ & 5:1 & 18 & 1&53& 10.6 \\%&-\\
10            &H$_2$O:$^{13}$CO$_2$ & 5:1 & 10&5 & 44& 8.8\\% & -\\
11          &H$_2$O:$^{13}$CO$_2$ & 5:1 & 10 & 1 &45& 9.0 \\%&-\\
12            &H$_2$O:$^{13}$CO$_2$ & 5:1 & 10 & 0.5&44& 8.8\\% &- \\
13           &H$_2$O:$^{13}$CO & 10:1 & 14 & 1 &  43& 4.3\\%&-  \\
14            &H$_2$O:$^{13}$CO & 10:1 & 25&1&47 & 4.7\\%& -\\
15           &H$_2$O:$^{13}$CO & 5:1 & 20 & 1 &24& 4.8\\%&-\\
16            &H$_2$O:$^{13}$CO & 2:1 & 13 & 1&9& 4.5\\%&- \\
17           &H$_2$O:$^{13}$CO & 1:1 & 17 & 1&4& 4.0\\%&- \\
18        & H$_2$O:CO$_2$:$^{13}$CO& 11:4:1 & 16 & 1&32/19& 12/2\\%&-/- \\ 
19        & H$_2$O:CO$_2$:$^{13}$CO& 20:1:1 & 30 & 1&92/96& 5/5\\%&-/- \\
%20        & CH$_4$ 							&		-			& 18	&	1	&	-	& -\\%&-	\\
%21		& H$_2$O:CH$_4$ 				&	5:1			&12	&	1	&	33& 6.6\\%&-\\

\hline                                   %inserts single line
\end{tabular}
\end{scriptsize}
\end{table}

\section{Experimental analysis}
\label{sec:res_exp}

\subsection{Complementarity of RAIRS and QMS}

\begin{figure}[!h]
\centering
	\resizebox{\hsize}{!}{\includegraphics{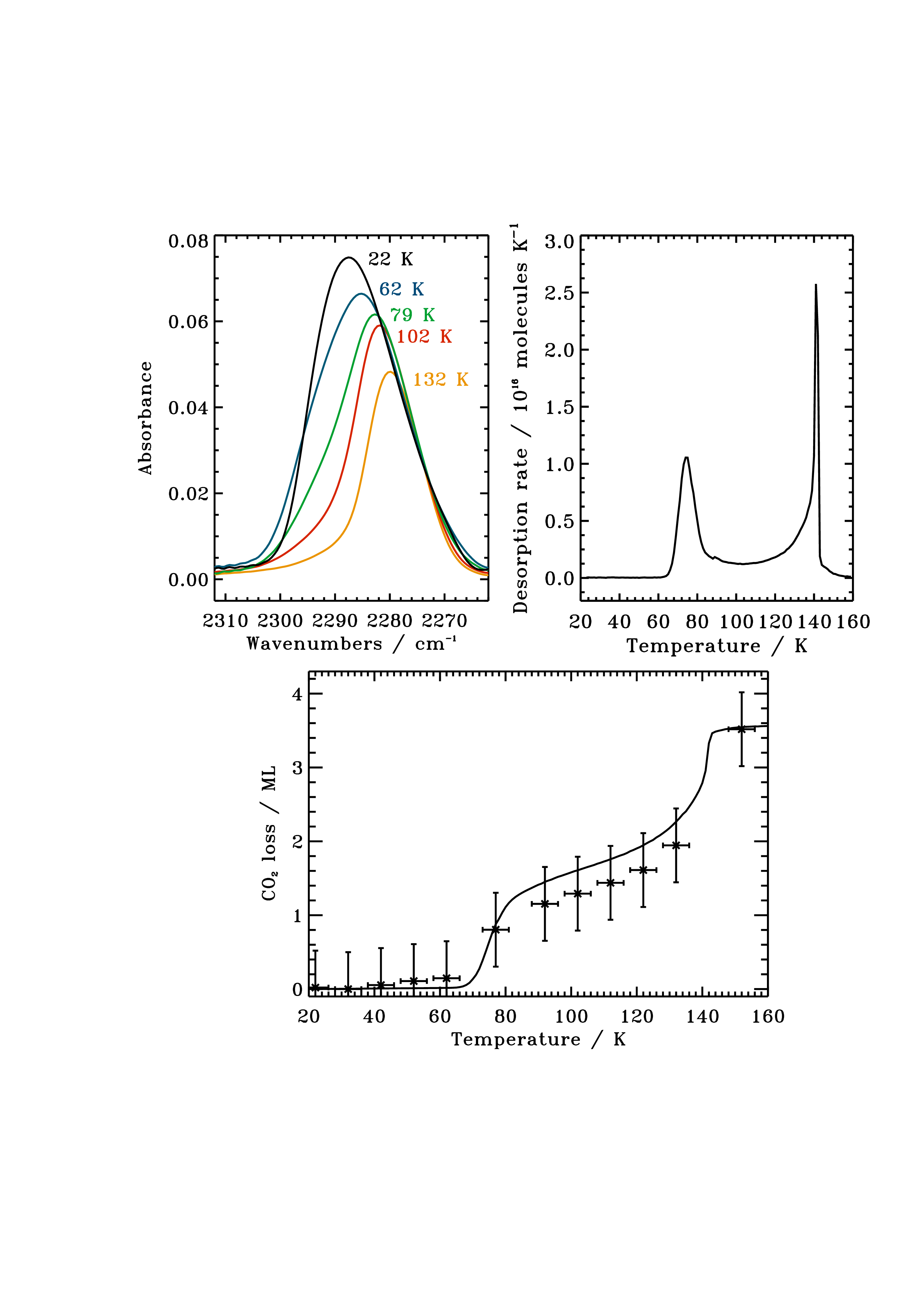}}
 \caption{The upper left panel presents the infrared CO$_2$ stretching features at specific temperatures during the warm-up of 18 ML of a H$_2$O:CO$_2$ 5:1 ice heated at 1 K.min$^{-1}$. The right upper panel presents the desorption rate of CO$_2$ for the same experiment obtained by mass spectrometry. The bottom panel shows the ice loss for this experiment obtained by infrared measurements (crosses) and mass spectrometry (solid line).
 }
 \label{qms_IR}
\end{figure}

Figure \ref{qms_IR} illustrates the agreement between desorption curves derived from QMS and RAIRS data for CO$_2$ in a 5:1 water-dominated H$_2$O:CO$_2$ ice, 18 ML thick and heated at 1~K.min$^{-1}$ rate (Exp. 11). The upper left panel in Fig. \ref{qms_IR} shows the CO$_2$ stretching band recorded at different temperatures during warm-up: after ice deposition at $22 \, \rm K$,  at 62~K where segregation is known to be efficient \citep{Oberg_09_c}, during the first ice desorption peak around 79~K, in the temperature interval between pure CO$_2$ desorption and H$_2$O desorption, and during desorption of the trapped CO$_2$. The right panel shows the desorption rate of CO$_2$ derived from the same experiment by mass spectrometry. The bottom panel presents the cumulative ice loss versus  temperature for this experiment, obtained both by integrating the CO$_2$ mass signal with respect to the temperature, and by integrating the CO$_2$ infrared signal recorded at specific temperatures. The error bars on the infrared data are due to variable ice band strengths with temperature and composition. Within these uncertainties the fractional ice loss curves derived by infrared and by mass spectrometry agree well; there seems to be only a small systematic offset for the 80~--~130~K range. This implies that %only 
the first RAIR spectrum of the ice after deposition can be used to derive quantitative results from the TPD experiments. 

Figure \ref{qms_IR} also shows that there is evidence for some ice loss between the two main desorption peaks. The cumulative QMS and infrared spectroscopy signals match each other at these intermediate temperatures,  which points to that the measurements trace actual ice desorption in between the pure ice desorption event and the desorption of trapped volatiles. The implications of this ice desorption process is discussed below, but it is important to note that this is not incorporated into the model framework and this may be a limitation to step-wise desorption models, whether using our parameterization or any of the previously published ones. Quantifying this process would require an additional set of experiments where the mass spectrometer is mounted closer to the substrate to allow for the detection of very low desorption rates.

\subsection{Desorption trends}

\begin{figure}[!h]
\centering
	\resizebox{\hsize}{!}{\includegraphics{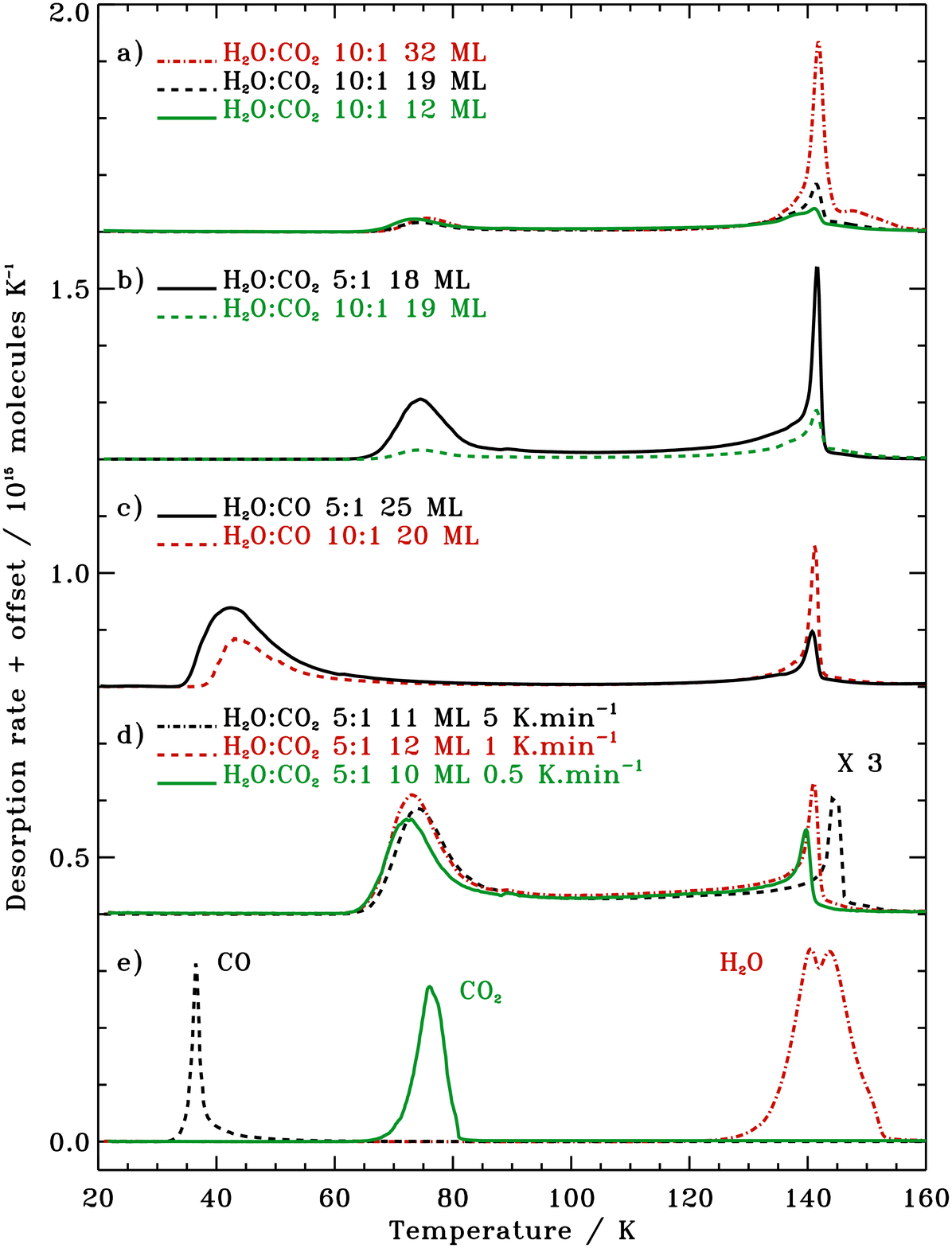}}
 \caption{Experimental CO and CO$_2$ desorption curves a-d) during warm up of ice mixtures (offset for visibility) together with pure CO, CO$_2$ and H$_2$O ice TPD curves e). The heating rate is 1 K.min$^{-1}$ except for when specified otherwise in a), the total ice thickness and mixing ratio are listed in for each experiment.
 }
 \label{tpd_exp}
\end{figure}

Figure \ref{tpd_exp} shows the desorption of CO$_2$ from H$_2$O:CO$_2$ ice mixtures of different thicknesses (a), with different CO$_2$ concentrations (b), and heated at different rates (d). In addition there are two CO TPD curves from H$_2$O:CO mixtures with different CO concentrations (c). For reference, Fig. \ref{tpd_exp}e) presents the TPD curves of pure CO, CO$_2$, and H$_2$O ice heated at 1 K min$^{-1}$. 
The fraction of trapped volatile is obtained by integrating the QMS signal for temperatures above $110\, \rm K$ and dividing it by the QMS signal integrated over the entire 20--160~K range. The chosen temperature of 110~K is well below the onset of the second desorption peak and the volatiles that desorbed during the first CO$_2$ or CO desorption peak are (almost) entirely pumped, though as discussed above there seems to be a low-level type of desorption occurring between the main desorption peaks. Whether due to finite pumping or actual desorption this results in a 10--20\% uncertainty in the determination of the trapped fraction, \textit{i.e.}, the choice of temperature integration limits affects the estimated amount of trapped ice by $<$ 20\%. The trapped percentage of volatiles in each experiment, defined with respect to the initial volatile ice content, is reported in the second last column of Table \ref{desorbexps}.The last column of Table \ref{desorbexps} presents the trapped abundance of volatiles species with respect to the initial H$_2$O abundance. This value is less variable compared to the trapped amount of CO/CO$_2$ with respect to the initial CO/CO$_2$ abundance presented in the preceding column. Both Table \ref{desorbexps} and Fig. \ref{tpd_exp} show that for CO$_2$ and CO the percentage of trapped volatile species in the H$_2$O ice is highly dependent on the experimental conditions; the CO$_2$ trapping fraction varies between 44 and 84\% with respect to the initial volatile content. In the following subsections, we report and discuss these dependencies.

	\subsubsection{Thickness dependency}
	
Figure \ref{tpd_exp}a presents the desorption of a H$_2$O:CO$_2$ 10:1 ice mixture for different initial ice thicknesses and shows that the amount of trapped CO$_2$ (desorption around 140~K) increases with ice thickness. In contrast the amount of CO$_2$ desorbing around 70~K is independent of ice thickness in the experimentally investigated range. This implies that only CO$_2$ molecules from the top part of the ice are available for desorption at the CO$_2$ desorption temperature. This can be explained by either a highly porous ice that allows CO$_2$ to "freely" desorb from the top layers or by diffusion from the top layers of the mantle phase to the surface.  In both cases the surface is eventually totally saturated by water molecules, trapping the rest of the volatiles in the ice mantle.
 
	\subsubsection{Mixing ratio dependency}

	The dependence of the volatile trapping with the mixture ratio is presented in Fig. \ref{tpd_exp}b) for H$_2$O:CO$_2$ and in Fig. \ref{tpd_exp}c)  for H$_2$O:CO ice mixtures. In both cases the trapped fraction decreases as the volatile to H$_2$O ratio increases. In other words, the amount of pores exposed to the surface or the diffusion length scale of volatiles in the ice must increase with increasing volatile concentration. A similar dependency was noted in \citet{Oberg_09_c} when measuring segregation in ices. Increased diffusion may either be due to a gradually looser binding environment in the volatile-rich ices or a break-down of H$_2$O ice structure in the presence of higher concentrations of volatiles. The thick H$_2$O:CO ice experiments (Exps. 14, 15 and 17)  with initial mixture ratios of 10:1, 5:1 and 1:1 show a continuous decline of the trapping fraction, which suggests that either the ice becomes continuously more porous or that the diffusion path length increases gradually with volatile concentration.
	
	\subsubsection{Molecular dependency}

Similarly to \cite{Sandford_90} and \cite{Collings_04} we find that the trapping efficiencies of the investigated CO$_2$ and CO in H$_2$O ice are radically different. When comparing Fig. \ref{tpd_exp}b) and c), it appears that  CO is much more mobile than CO$_2$ in the H$_2$O ice as demonstrated by the higher trapping fraction of CO$_2$ compared to CO in similar H$_2$O ice mixtures. \cite{Sandford_90} explained this difference from a combination of different binding energies of CO and CO$_2$ in H$_2$O ice due to molecular size, shape and electronic differences. These binding energies may equally affect the probability of escaping through an ice pore or diffusing through the bulk of the ice.

	\subsubsection{Heating rate dependency}
	
	In Fig. \ref{tpd_exp}d), the heating rate of the ice is varied for a H$_2$O:CO$_2$ 5:1 ice of $10-12\, \rm ML$ between $0.5, \,1 \rm \, and \,5 \, \rm K.min^{-1}$. This does not appreciably affect the trapping efficiency of CO$_2$ in the H$_2$O ice and implies that the process responsible for exchanges between the surface and the mantle is fast compared to the experimental warm-up time. If this was not the case, a lower heating rate would have resulted in a smaller amount of trapped CO$_2$, since a slower heating means more time for the migration of mantle molecules to the surface. 

The lack of a heating rate dependency on the trapped amount of volatiles (same trapped percentage in Exp. 10, 11 and 12) also implies that there is a rather sharp boundary between the molecules in the upper layers that can diffuse to the surface (whether through pores or bulk diffusion) and molecules deeper in the ice that cannot. Even if the volatiles deep in the ice can diffuse within the ice mantle, diffusion 'upwards' must quickly become slow as the surface layers saturate with H$_2$O molecules or alternatively all accessible pores have been emptied. This explains that the amount of desorbing CO$_2$ molecules at low temperatures is thickness independent (the H$_2$O `ice cap' will become impenetrable after a certain amount of CO$_2$ molecules have desorbed) and that entrapment efficiencies are unaffected by lower heating rates.

\subsection{Tertiary mixtures}

\begin{figure}[t]
\centering
	\resizebox{\hsize}{!}{\includegraphics{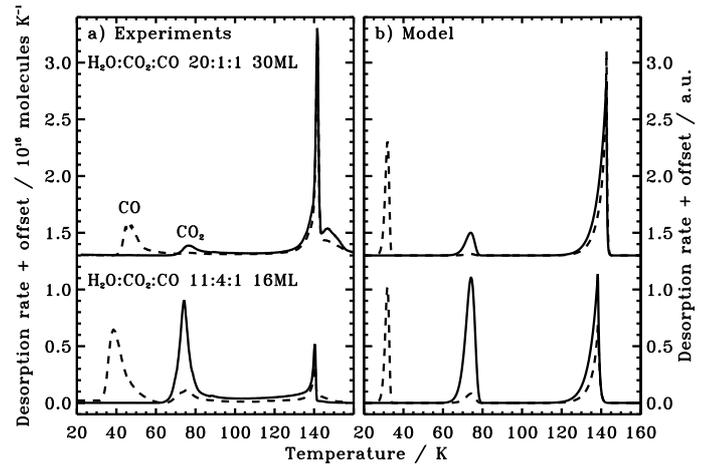}}
 \caption{a: Desorption rate of CO$_2$ (solid line) and CO (dashed line) from two tertiary water dominated ice mixtures (Exps. 18 and 19) with a 1 K.min$^{-1}$ heating rate. - b: Implemented three-phase model desorption rate for the same eperiments.
 }
 \label{qms_tert}
\end{figure}

The desorption rates for the tertiary mixtures (Exps. 18 and 19) are presented in Fig. \ref{qms_tert}a). Similarly to the binary experiments, the trapping of volatiles is more efficient for a lower volatile to H$_2$O ratio. The desorption curves for CO$_2$ are not affected by the presence of CO. Thus the CO-CO$_2$ interaction does not have a significant impact on the amount of CO$_2$ trapped within the water matrix. Overall the TPD curves resemble the addition of desorption curves from two separate binary mixtures, except for a small fraction of CO that desorbs with CO$_2$ in the H$_2$O:CO$_2$:CO 11:4:1 mixture (lower panel in Fig. \ref{qms_tert}). It is unclear whether this desorption is due to a co-desorption of CO with CO$_2$ or to a release of CO that has been trapped under a barrier of CO$_2$ surface molecules. The observed similarity supports the use of binary ice mixtures as templates to study diffusion and desorption even though they are not directly representative of interstellar ice mixtures.

When comparing the desorption of CO from a tertiary and a binary ice mixture with the same H$_2$O:CO ratio and ice thickness, it appears that less CO is trapped in the tertiary mixture. This may be due to ice structure changes, as discussed above, or to shielding of CO from the sticky water molecules by CO$_2$ molecules (the CO-CO$_2$ bond is weaker than the H$_2$O-CO one), lowering the CO diffusion barrier.
The CO$_2$ and CO desorption curves from the dilute tertiary mixture both contain a small additional peak around 150~K, only seen elsewhere in the thickest H$_2$O:CO$_2$ 10:1 ice experiment. A similar double peak was noted in the 20:1 desorption experiments of \citet{Collings_04}.

\subsection{Ice diffusion mechanisms: pore versus bulk diffusion}

The main mechanism behind diffusion in the ice mantle is not known and may differ between different ices. Most previous studies have focused on diffusion in cracks and pores and pore collapse has been introduced to explain ice trapping. The observation that both CO and CO$_2$ become trapped even though they partly desorb at their, very different, pure ice desorption temperatures is difficult to reconcile with pore collapse as the main trapping mechanism, however. That is, it would imply efficient H$_2$O pore collapse both at $\sim$30~K and $\sim$70~K.

Even if pore collapse does not provide a complete explanation of why ices become trapped, some kind of internal surface hopping may explain why molecules can diffuse out of the ice. In this scenario, the CO$_2$ desorption ice thickness dependency is due to that the pores and cracks that are open to the surface only go down to a certain depth, in this case a few ML for H$_2$O:CO$_2$ 5:1. The different CO and CO$_2$ trapping efficiencies may then be either due to different ice structures or to CO desorbing easier through pores compared to CO$_2$. Pore diffusion may thus be consistent with these particular experimental results, but ice diffusion is present also in other ices that are known to be quite compact, e.g. CO ice \citep{Bisschop_06}. While it is possible that diffusion occurs through completely different mechanisms in different ices, the concept of ice bulk diffusion has the prospect of approximating mixing and de-mixing processes in all kinds of ices, regardless of structure.

In the bulk diffusion scenario, an amorphous ice is viewed like a very viscous liquid, whose viscosity decreases with the volatility of the ice molecules. In \citet{Oberg_09_c} this was modeled as molecules swapping places with a barrier significantly higher than surface hopping. Volatile molecules will tend to swap their way towards the surface because it is energetically favorable to have the molecules that form weaker bonds in the surface layer (where fewer bonds can be made due to the ice--vacuum interface). Trapping is explained by that when volatile molecules diffuse from the top mantle layers to the surface and desorb, the top mantle layers become saturated with H$_2$O and therefore viscous enough to be impenetrable. The low desoption rate of volatiles between the volatile and H$_2$O ice desorption peaks would however suggest that under some experimental conditions, small amounts of volatiles can escape through this H$_2$O barrier. More experiments are required to test under which conditions this is a reasonable approximation. The model presented below is an attempt to include the most important features of this concept while still keeping the number of parameters low.

\section{Model parametrization and performance}
 \label{sec:mod}

\subsection{Parametrization}

\begin{figure}[t]
\centering
	\resizebox{\hsize}{!}{\includegraphics{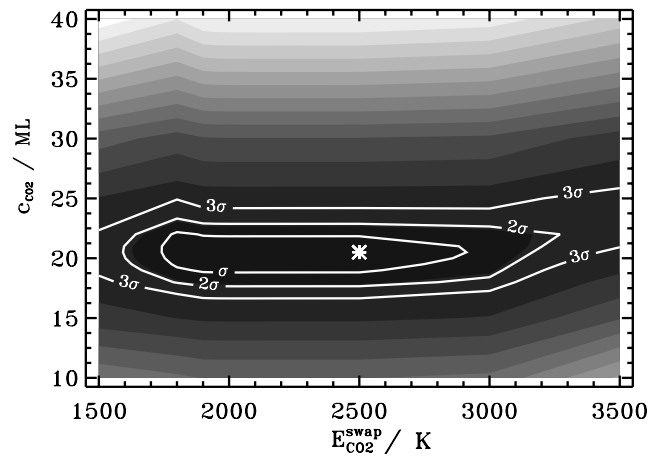}}
 \caption{ $\chi^2$ contour plot for fitting the model parameters $\epsilon_{\rm H_2O-CO_2}^{\rm swap}$ (parameterizing the ability of CO$_2$ to switch position with a H$_2$O molecule) and $c_{\rm CO_2}$ (parameterizing the CO$_2$ ice thickness where swapping is fast compared to the investigated heating rates) using the experimentally determined amount of CO$_2$ trapped in a binary H$_2$O:CO$_2$ ice (Exps. 4 to 12). These two parameters regulate the distribution of the volatile molecules in the gas, surface and mantle according to Eq. \ref{eqr_8} and \ref{eq_energy} of Section \ref{sec:model}.
 }
 \label{chisquare}
\end{figure}

The pure ice desorption energies are derived from fitting the three-phase model to the experimental pure ice desorption curves with the results: $E_{\rm H_2O} = 4400 \rm \, K$, $E_{\rm CO_2} = 2440 \rm \, K$ and $E_{\rm CO} = 1010 \rm \, K$. The H$_2$O value is lower than the one found in \citet{Fraser_01}. This discrepancy is probably due to a combination of that we use a single experiment, do not include the pumping speed and fix the pre-exponential factor to $\nu = 10^{12} s^{-1}$  for every species here. The value in \citet{Fraser_01} should thus still be used when modeling the absolute desorption temperature. For the purpose of parameterizing the desorption fractions we prefer our value for the sake of consistency. The remaining model parameters $\epsilon_{\rm H_2O- \textit{i}}^{\rm swap}$ and $c_i$ are obtained separately for  CO$_2$ and CO. This is done through a $\chi^2$ analysis, where trapping fractions from the model are compared to those from the binary experiments for a grid of $\epsilon_{\rm H_2O- \textit{i}}^{\rm swap}$ and $c_i$ values. The minimum $\chi^2$ value for CO$_2$ is obtained for $\epsilon_{\rm CO_2-H_2O}^{\rm swap}$=2500 K, $c_{\rm CO_2}$=20.5 ML, but  the swapping energy for H$_2$O:CO$_2$, which is linked to the ability of CO$_2$ to swap with H$_2$O within the solid phase, is not well-constrained between 1600-3200~K (Fig. \ref{chisquare}). In contrast, the parameter $c_{\rm CO_2}$ related to available amount of CO$_2$ from the ice mantle that can migrate to the surface, is well constrained which suggests that even at laboratory time scales mixture desorption is mainly governed by how large a part of the mantle is eligible for swapping with the surface, rather than the swapping barriers.

 The CO experiments can be fitted with $\epsilon_{\rm CO-H_2O}^{\rm swap}$= 960 K and $c_{\rm CO}$= 80 ML, but these are based on only a few experiments and the inequalities $ \epsilon_{\rm H_2O- CO}^{\rm swap}<\epsilon_{\rm H_2O- CO_2}^{\rm swap} $ and $c_{\rm CO}>c_{\rm CO_2}$ are alone well constrained.

\subsection{Model performance}
\subsubsection{Desorption trends modeling}

\begin{figure}[t]
\centering
	\resizebox{\hsize}{!}{\includegraphics{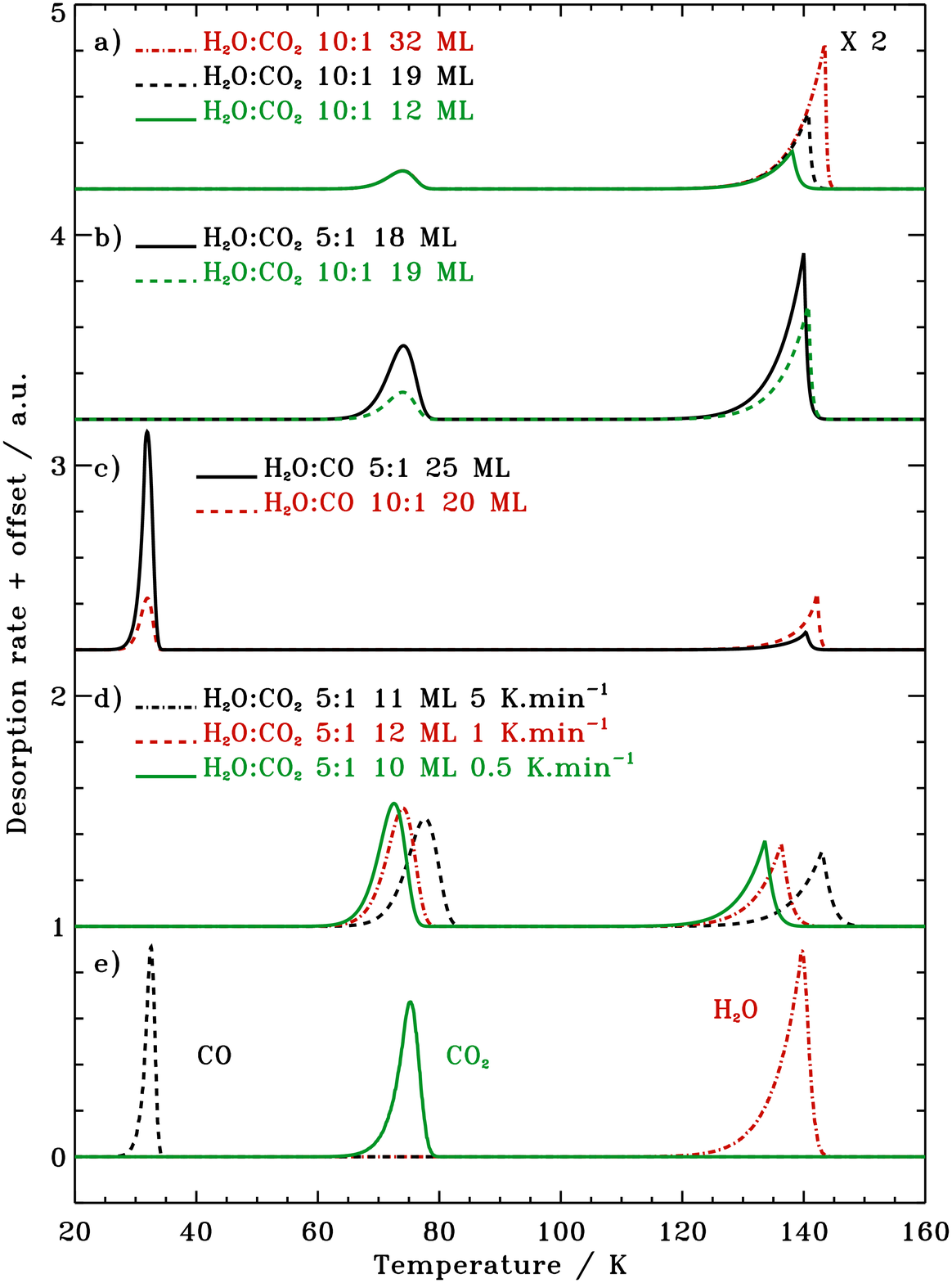}}
 \caption{Simulated CO$_2$ and CO TPD curves from H$_2$O:CO$_2$ and H$_2$O:CO mixtures for different thicknesses, ratios, and heating rates a-d). Panel e) presents the simulated desorption of pure H$_2$O, CO$_2$ and CO ice. This figure connects the model outputs to the experiments shown in Fig. \ref{tpd_exp}. }
 
 \label{tpd_sim}
\end{figure} 

Figure \ref{tpd_sim} shows the simulations of the binary mixture desorption using the optimized model parameters from the previous section. Generally the qualitative agreement is good and the model captures the trends that were observed experimentally. The exact shapes of the modeled and experimental desorption curves differ for several reasons. First, the model does not take into account the range of environments from which the molecules desorb, \textit{e.g.}, the H$_2$O/volatile fraction changes during the desorption process and even a pure ice has a range of different binding sites. This affects both the position and the width of the peak. Second, the model does not consider the different water ice structures present at different temperatures; water ice crystallizes around 140~K (see the desorption peaks for water in Fig.~\ref{tpd_exp}e), which may affect the shape and position of the second desorption peak of the volatiles. Finally, the model does not include finite pumping speeds, which results in %artificial desorption 
abundance tails in the experimental curves.
While these effects may all be important under special circumstances the aim of the extended three-phase model is not to reproduce the experimental results perfectly. Rather, the goal is to capture the main characteristics of ice mixture desorption. %Below we discuss why the experimental trends are well reproduced by the model.

The increase of the trapped amount of volatiles with ice thickness and the experimental observation that the same amount of molecules desorbs around the volatile pure desorption temperature, regardless of the ice thickness, are reproduced in the model because diffusion between the mantle and surface is only allowed from a fraction of the mantle, $f_i$ (Eq. \ref{eqr_8} in Section \ref{sec:model}) which depends on the kind of volatile and mixing ratio with water. This fraction is %set to be 
independent of the ice thickness. Thus the same amount of volatile molecules migrates to the surface regardless of ice thickness, followed by saturation of the surface phase by water molecules. The rest of the volatiles is trapped in the mantle until H$_2$O desorption, thus the trapped fraction depends on the ice thickness.

The observed concentration effect on the trapping efficiency is reproduced by the model because the fraction of volatiles migrating to the surface depends on the mixing ratio of the volatile with respect to water, $x_{i}^{\rm ini}$. The lower the concentration of volatiles in the ice is, the smaller the fraction of volatile molecules make it to the surface and the more become trapped. The higher mobility of CO compared to CO$_2$ is also reproduced by the model as the molecular paramater $c_{\rm CO_2}$ is lower than $c_{\rm CO}$. Thus more volatiles are able to diffuse to the surface in the case of H$_2$O:CO mixtures.\

Experimentally, a low desorption rate is sometimes observed between the volatile and H$_2$O desorption temperature. In the model, the diffusion barrier energy is low enough that the diffusion process is complete before desorption of the volatile takes place and therefore there is no desorption between the pure and H$_2$O desorption peaks. Such a low diffusion energy barrier is needed  to reproduce that the trapping efficiency is insensitive to the heating rate (within the explored heating rate range). There is probably a second diffusion process at play at these intermediate temperatures, which cannot be reproduced by the current, simple parameterization.  \

\subsubsection{Quantitative agreement}

 \begin{figure}[t]
\centering
	\resizebox{\hsize}{!}{\includegraphics{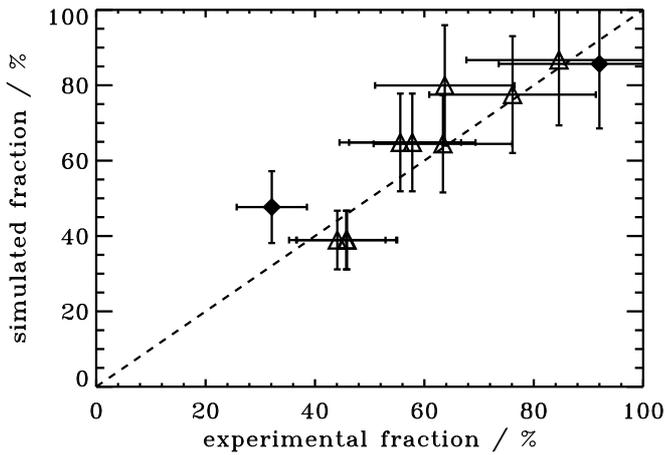}}
 \caption{Percentage of CO$_2$ experimentally trapped vs the simulated one. The triangles are the trapped percentages obtained from the binary mixture experiments. The diamonds are the model predictions for the two tertiary mixtures. The one-to-one ideal relation is plotted as a dashed line. }
 
 \label{tpd_agree}
\end{figure} 

 Figure \ref{tpd_agree} compares the volatile trapping fractions obtained by the optimized H$_2$O:CO$_2$ model to those found experimentally. The error bars include the uncertainties due to the choice of the temperature from which we integrate the second QMS peak and are between 10--20\% for the different experimental data points. The error bars on the model results originate from the uncertainties in the input ice thicknesses, mixing ratios and binding energies and these were obtained by varying the model input values within the experimental uncertainty ranges and then comparing the model results. In general, the uncertainty in the mixing ratio has the largest effect, resulting in model prediction uncertainties of $\sim$15\%. We conclude that CO$_2$ desorption from binary mixtures is quantitatively described by the model.  

\subsubsection{Predictive power} 
 
 Figure \ref{qms_tert}b shows the output of the model for the H$_2$O:CO$_2$:CO tertiary ice mixture experiments (Exp. 18 and 19). These experiments were not used to constrain the model and are as a test of its predictive power. The concentration dependency in these experiments is reproduced by the model; an increase in the concentration of volatiles leads to a decrease in the trapping fraction. 
 
 For the higher concentration mixture, H$_2$O:CO$_2$:CO = 11:4:1 (Fig. \ref{qms_tert}b, bottom panel), the model gives a CO desorption peak around 70 K (corresponding to the pure CO$_2$ desorption temperature), peak that was also experimentally observed (Fig. \ref{qms_tert}a, bottom panel). In the model, this peak results from the formation of free desorption sites on the surface due to desorption of surface CO$_2$. CO molecules that are mixed with water migrate to and desorb from the surface easily since the swapping and binding energies are very low compared to the CO$_2$ values.

In addition to reproducing these qualitative trends for tertiary mixtures, the three-phase model also treats correctly the desorption order of both CO$_2$ desorption peaks. Finally Fig. \ref{tpd_agree} shows that the model provides a reasonable quantitative agreement between the predicted and experimentally determined amounts of CO$_2$ trapped in the tertiary ice mixtures (black diamonds). This is very promising for extending this proof-of-concept model to more species and more complex mixtures. 

\section{Astrophysical implications}
 \label{sec:astro}
Trapping of volatiles in H$_2$O ice is a crucial parameter when predicting the chemical evolution during star and planet formation \citep{Viti_04}. The modified three-phase desorption model is used here to test the effect of different initial ice compositions and ice thicknesses on ice mixture desorption. Ultimately the three-phase model should, however, be integrated in a protostellar collapse model to simulate the ice desorption accurately during star formation. The prime advantage of the three-phase model, as initially introduced by \cite{Hasegawa_93} and extended here, is that it can treat surface and ice chemistry correctly, since it differentiates between surface molecules that can react with gas phase molecules, and mantle molecules that are protected from further processing.

 \begin{figure}[t]
\centering
	\resizebox{\hsize}{!}{\includegraphics{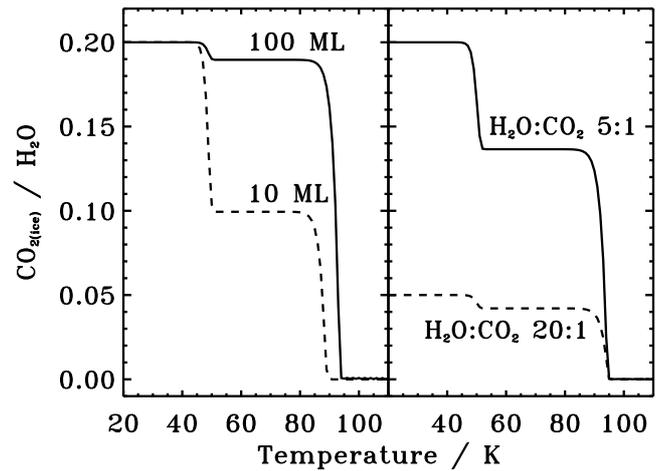}}
 \caption{The amount of CO$_2$ ice during ice warm-up at 1 K per 100 years according to the three-phase model, assuming two different initial H$_2$O:CO$_2$ 5:1 ice mixture thicknesses (left panel) and two different 20 ML ice mixing ratios (right pannel).}
 
 \label{astro_fig}
\end{figure} 

 \begin{figure}[t]
\centering
	\resizebox{\hsize}{!}{\includegraphics{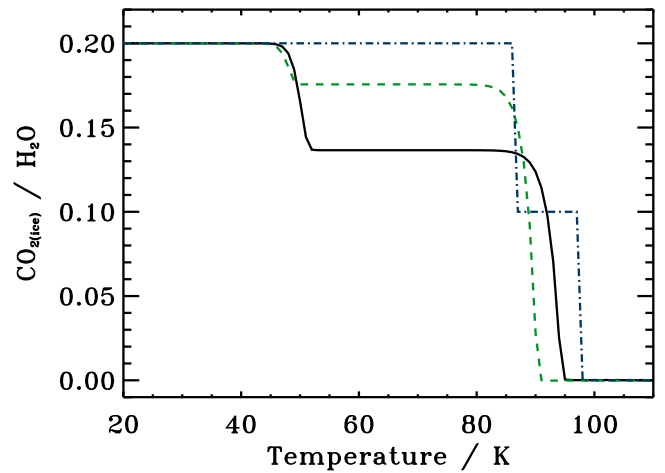}}
 \caption{Amount of CO$_2$ ice during ice warm-up for a H$_2$O:CO$_2$ 5:1 ice 20 ML thick simulated by different models : the implemented three-phase model described here (black solid line), the original three-phase model by \cite{Hasegawa_93} (green dashed line), the \cite{Viti_04} model (blue dash-dotted line). A heating rate of 1~K per century is used in the two first models and desorption around a 5 solar masses protostar is presented from the \cite{Viti_04} model case.}
 
 \label{astro_fig_2}
\end{figure} 

Figure \ref{astro_fig} shows the amount of CO$_2$ ice with respect to the original H$_2$O ice abundance as a function of temperature for different ice thicknesses and mixing ratios heated at 1~K per 100 years, typical for infall of material during protostar formation \citep{Jorgensen_05_b}. 

The percentage of CO$_2$ entrapment in a diluted water ice is significantly affected by the initial ice thickness and mixing ratio;  50\% of the initial CO$_2$ abundance is trapped in a 10 ML ice and 95\% in a 100 ML ice. A similarly dramatic difference is seen when assuming different initial ice mixtures: 64\% of the CO$_2$ stays trapped in the 5:1 ice and the fraction increases to 84\% for the 20:1 ice.

The treatment of these trapping dependencies is one of the key strengths of the extended three-phase desorption model presented here. 
Figure \ref{astro_fig_2} compares CO$_2$ ice desorption from a H$_2$O:CO$_2$ ice using the extended three-phase model, the original three-phase model by \cite{Hasegawa_93}, and the \cite{Viti_04} astrochemical network. Assuming a 20 ML thick H$_2$O:CO$_2$ 5:1 ice heated at 1~K per century, our model predicts that 64 \% of the initial CO$_2$ will be trapped by the water ice, while the model by \cite{Hasegawa_93} predicts an 80 \% trapping amount. This difference originates from the lack of mantle-surface diffusion in \cite{Hasegawa_93}. Its implementation is clearly important to correctly treat trapping of volatiles and to account for segregation observed around protostars \citep{Ehrenfreund_98, Pontoppidan_08}. 

 All the CO$_2$ molecules are predicted to be trapped by the water ice when simulating H$_2$O:CO$_2$ ice desorption with the \cite{Viti_04} model. The \cite{Viti_04} model assumes a different heating rate compared to the one used for the two three-phase models, but this only affects the desorption temperatures and does not affect the volatiles trapping fractions. Instead, the high trapping fraction is due to the fact that the model was parametrized based on a desorption experiment performed for a H$_2$O:CO$_2$ ice with a ratio of 20:1, which differs from the 5:1 -- 4:1 ratio found in dense molecular clouds and protostellar envelopes \citep{Knez_05,Pontoppidan_08}. In the case of a H$_2$O:CO$_2$ 20:1 ices, our model outputs agree well with 100 \% trapping fraction used by \cite{Viti_04}, since we find that more than 95 \% of the CO$_2$ is trapped by the water ice for 10 -- 100 ML thick ices. 

These different model predictions demonstrate the need for systematic laboratory studies when modelling ice desorption, since ice properties, such as ice thickness and mixing ratio, affect the desorption process. Even when using desorption step functions, the size of the step cannot be accurately decided from a single experiment. Rather the investment of multiple experiments are needed, together with their efficient parameterization, to obtain versatile models of ice desorption for arbitrary initial conditions. Already for binary ice mixtures, this results in large experimental data sets. It is therefore reassuring that using binary mixtures as templates for more complex ice mixtures results in approximately the correct trapping predictions.

\section{Conclusions}

Desorption from H$_2$O-rich ice mixtures is complex in that the amount of trapped ice depends not only on the species involved, but also on the mixture ratio and the ice thickness ; there is no constant fraction of volatile species trapped in a H$_2$O ice. This complex behavior can be reproduced by extending the three-phase model introduced by \citet{Hasegawa_93}. 

Using the H$_2$O:CO$_2$ ice system as a case study, we showed that a three-phase model that includes mantle-surface diffusion can reproduce the amount of trapped ice quantitatively in a range of binary ice mixtures. The appropriate input parameters for the H$_2$O:CO$_2$ system are a swapping energy  $\epsilon^{\rm swap}_{\rm H_2O - CO_2}$=2250 K and a molecular parameter $c_{\rm CO_2}$=20.5 ML, which describes from which ice depth diffusion to the surface can occur. 

In the model, the different CO$_2$ and CO behavior can only be reproduced if $ \epsilon_{\rm H_2O- CO}^{\rm swap}<\epsilon_{\rm H_2O- CO_2}^{\rm swap} $ and $c_{\rm CO}>c_{\rm CO_2}$. This suggests that diffusion/molecule swapping in H$_2$O-rich ices depends equally on the breaking of the H$_2$O-volatile bond (parametrized here by the swapping barrier height) and on the mass/volume of the diffusing volatile (parameterized here by the ice thickness that the molecule can diffuse through). The experimental trends found for H$_2$O:CO and H$_2$O:CO$_2$:CO ice mixture desorption are consistent with the H$_2$O:CO$_2$ trends, which suggests that the three-phase model is generally appropriate to model ice mixture desorption. 

However the ice desorption process is implemented in astrochemical models, this study demonstrates that it is vital to understand how ice mixture desorption depends on the ice characteristics. The extended three-phase model naturally treats ice desorption with the right kinetic order and reproduces volatile entrapment. Its use in astrochemical networks for grain-gas interactions should improve the predictions of gas-phase and grain-surface species abundances in astrophysical environments.

\section*{Acknowledgements}

This work has greatly benefitted from discussions with Ewine van Dishoeck and comments from an anonymous referee. Funding is provided by NOVA, the Netherlands Research School for Astronomy. Support for K.~I.~\"O. is provided by NASA through Hubble Fellowship grant  awarded by the Space Telescope Science Institute, which is operated by the Association of Universities for Research in Astronomy, Inc., for NASA, under contract NAS 5-26555. 

 	\bibliography{mybib}
 	
\end{document}